# Classification of IDS Alerts with Data Mining Techniques

Hany Nashat Gabra
Computer and Systems Engineering Department, Ain Shams University, Cairo, Egypt.
hanynashat@hotmail.com

Dr. Ayman M. Bahaa-Eldin
Computer and Systems Engineering Department, Ain Shams University, Cairo, Egypt.
ayman.bahaa@eng.asu.edu.eg

Prof.Huda Korashy
Computer and Systems Engineering Department, Ain Shams University, Cairo, Egypt.
hoda.korashy@eng.asu.edu.eg

## ABSTRACT

Intrusion detection systems (IDSs) have become a widely used measure for security, but we still have a problem on those systems results which includes many irrelevant alerts, so we will propose a data mining based method for classification to distinguish serious alerts and irrelevant one with the performance of 99.9 % in comparison with the other recent data mining methods which have reached the performance of 97%. Also we create a list of alerts sorted by alert's importance to minimize the human interventions.

**Keyword**: Intrusion Detection, Data Mining, Frequent Pattern, Frequent Itemset, support

## 1. INTRODUCTION

Although the IDS is a security measure for network monitoring and protection but unfortunately the IDSs are known to generate large amounts of alerts, with many of them being either false positives or of low importance. Which is hard for human to find alerts that need more attention, in order to tackle this difficulty; we proposed an IDS alert classification and alert ranking method based on an enhanced data mining technique.

An IDS sensor can generate thousands of alerts in a day [1, 2]. Furthermore, often vast majority of the alerts are false positives or of low importance [3,1]. As mentioned that it is usual to receive thousands of alerts from a single network IDS sensor per day, with more than 90% of the alerts being irrelevant [16, 20, 8] so an IDS alert log's analysis techniques are often used to distinguish the important IDS alerts from irrelevant events. Many approaches have been suggested like machine learning [4], time series modeling [8, 5], the use of the control charts [6], etc. During the last 10 years, data mining based methods have also been proposed in many researches papers [16, 20, 7, 3, 9, 10]. With these methods, IDS alert logs that recently logged are mined for knowledge that is used for the creation of event filtering for future IDS alerts.

However, existing methods are inherently automated or semi-automated but they assume that a human expert interprets detected knowledge and creates event filtering and correlation rules by hand. In the proposed technique we classified the alerts based on a data mining technique that has been enhanced for the IDS systems to distinguish serious alerts and irrelevant one. The result showed that the performance has been enhanced as we reduced the number of alerts to 99.9 % in comparable with the performance of other recent techniques which have reduced the number of alerts by 74-97% [16, 2, 9, 10, 1].

## 2. RELATED WORK

Data mining techniques first used for knowledge discovery from telecommunication event logs more than a decade ago [13]. In the context of IDS alerts mining, a number of techniques have been suggested. Clifton and Gengo [7] have investigated the detection of frequent alert sequences and enhanced by Ferenc [11], Walter A. Kosters and Wim Pijls [12], to use this knowledge for creating IDS alert filters. Long et al. [3] suggested a clustering algorithm for distinguishing Snort IDS true alerts from false positive alerts. Julisch and Dacier [16] also proposed a conceptual clustering technique for IDS alert logs as well, so that clusters correspond to alert descriptions, and a human experience can be used for developing filtering and correlation rules for future IDS alerts. Many of other approaches have been suggested like machine learning [4], time series modeling [8, 5], the use of the control charts [6], etc. During the last 10 years, data mining based methods have also been proposed in many researches papers [16, 20, 7, 3, 9, 10].

## 3. THE ALERT CLASSIFIER

A frequent pattern based outlier detection based method for discovering knowledge from the IDS alert logs and creating alert classifiers from this knowledge in a semi-automated manner has been used on our proposed technique. This method is motivated by the main drawbacks of previously proposed data mining approaches and its automation that considered mistakenly serous alerts as an irrelevant or fails positive. On 2009 Risto Vaarandi [1] proposed an automated technique and had created a main drawback as mentioned of classifying true attacks as a false positive.
Some researchers have observed the that the most alerts are triggered by only a few signatures [1] and in [8] it is reported that 68% of the alerts were produced by five signatures, while according to [5] seven signatures produced 77% of the alerts.
Also it has been noticed that if a signature has triggered many alerts over a longer periods of time, it is most properly will do so in the future. When we investigated routine alerts from IDS sensors, we found that they are either false positives or events of low importance. A prominent alert of low importance is related to MS Slammer Sapphire worm which infected many computers around the world in 2003. Despite vast majority of these computers were cleaned and patched, there are still infected nodes around that are constantly scanning the Internet for victims. Although this malicious network traffic triggers many alerts, they represent a very frequent and well-known attack that doesn't pose a threat to properly maintained systems.
We believe that the identification of such routines and alerts ranking is important. First, it helps to save human effort that is spent for editing alert filters. Therefore, security analysts will have more time for reviewing important alerts which don't match routine alert patterns and thus deserve closer investigation. Second, since most IDS alerts are routine events, there will be much less important alerts to investigate than in the original IDS log.

Third, we improve the detection result reliability of the automated method that has been used by Risto Vaarandi [1, 21]. Risto Vaarandi used LogHound to employ a frequent itemset mining algorithm for discovering frequent patterns from event logs. We proposed a new technique in order to detect the true attacks directly by ranking all alerts according to

Classification of IDS Alerts & Ranking using Data Mining Techniques

their impotency and save human effort. The new outlier detection methodology have applied to rearrange the snort output according to the alert importance so security analysts will have more time for reviewing the important alerts first then to investigate the less important one.

Outlier detection is a new data mining technique which has absorbed many attentions recently. It is able to identify abnormal data (called outlier) in large datasets.

Frequent pattern-based outlier detection is presented by Zengyou He et al. in [19]. This method is built on the following truth: given a set of supermarket transactions, where each transaction is a set of literals (called items).

Frequent itemsets (also called frequent patterns) are those combinations of items that have transaction support above predefined minimum support (support means percentage of transactions containing these itemsets). If a transaction contains many frequent itemsets, it means that this transaction is unlikely to be an outlier because it possesses the "common features". Accordingly, by employing the Frequent pattern-based outlier detection based method the cost wasted for processing false positives will be reduced and the reaction to attacks will be more rapid.

## 4. Mining Frequent Patterns

Mining frequent itemsets from a database has been solved largely by algorithms that are Apriori based and those that are pattern-tree growth techniques. Algorithms for mining of all existing techniques do not include generating frequent patterns for each transaction or by anther meaning do not show the records where they occurred with their transaction ids.

For many applications, just producing the frequent patterns without linking them to the specific transactions they occurred in may not be adequate. None of the frequent itemset mining algorithms considers mining frequent patterns with their transactions.

| Alerts | Item |
|---|---|
| Alert1 | 1 3 4 |
| Alert2 | 2 3 5 |
| Alert3 | 1 2 3 5 |
| Alert4 | 2 5 |

**Table 1: Example Alerts/items Data set Records**

Assume a data set which contains alert records generated by an IDS system in Table 1 where the set of items I = {1, 2, 3, 4, 5} and the set of Alerts = {Alert1, Alert2, Alert3, Alert4}.
Mining all alerts that have similar frequent itemset at minimum support of 50% would require generating frequent itemsets with the alerts in the format [< itemset > Alerts-list] that allows mining more informative large itemsets as:
L = { [< 1 > Alert1Alert3], [< 2 > Alert 2 Alert 3 Alert 4],
< 3 > Alert 1 Alert 2 Alert 3], [< 5 > Alert 2D3 Alert 4], [< 1, 2 > Alert 2 Alert 3 Alert 4],[< 1, 3 > Alert 1 Alert 3],
[< 2, 3 > Alert 2 Alert 3], [< 2, 5 > Alert 2 Alert 3 Alert 5], [< 3, 5 > Alert 2 Alert 3], [< 2, 3, 5 > Alert 2 Alert 3]}.
We proposed the AlertFp algorithm for mining frequent patterns with the Alerts where they occurred. Mining Fps with Alerts on an IDS log is an important goal of this algorithm where we are linking all frequent patterns to the data set records or alert's transactions where they came from. Then count the number of frequent patter founded on each transaction.
Finally all transactions in the dataset re-sorted according to the number of the related frequent patters.
AlertFp algorithm represents each frequent k-pattern as form < $F_{k1}$, $Alert1_{k1}$, $Alert2_{k1}$, . . . , $Alertm_{k1}$>, where $F_{k1}$ is the first frequent k-pattern, and $Alertm_{k1}$ is the mth Alert of the first frequent k-pattern.
For example, given Table1 and mini support of 50% (pattern exists on 50% of the whole alerts at least / 2 alerts at least), the list of frequent 1-itemsets is F1 ={< 1, Alert1,Alert3 >, < 2,Alert2, Alert3,Alert4 >, < 3, Alert1, Alert2, Alert3 >, < 5, Alert2, Alert3, Alert4 >}.
This implies as well that the candidate 1-itemsets listed by this technique is in the same form as:
C1 ={< P1, $Alert1_1$, $Alert2_1$, . . . , $Alertm_1$ >}, where P1 is the first candidate 1-pattern, and $Alertm_1$ is the mth Alert of the first candidate 1-pattern.
For our example data set, the candidate 1-pattern is given as C1 ={< 1, Alert1,Alert3 >, < 2, Alert2, Alert3, Alert4 >, < 3, Alert1, Alert2, Alert3 >, < 4, Alert1 >, < 5, Alert2, Alert3, Alert4 >}.
Thus, with this AlertFp technique, the data set is scanned only once to obtain the candidate 1-itemsets with a list of their Alerts. The Alerts of each candidate pattern is implemented. Then, the count of each candidate pattern's Alerts is equivalent to the support of the pattern. The pattern or the itemsets having support less than the mini support are excluded from the frequent 1-itemsetAlertFP leading to the itemset <4, Alert1> being deleted from the C1 list to get F1 as we predefined the mini support as 50%.
AlertFp algorithm applies the Apriori function, which works on two components of the itemsets consisting of the itemset part and the alert part and obtaining higher order frequent itemsets for their supports.
After applying the frequent pattern mining algorithm to past IDS alert logs (AlertFp), in order to discover patterns that describe redundant alerts.
Alert weight is measured by calculating Frequent Pattern Outlier Factor (FPOF) for each alert's transaction.

$$FPOP\ (t) = \frac{\sum_{(D, minisupport)} support\ (X)\ X \subseteq t,\ X \in FPS}{\|FPS(D, minisuport\ t)\|}$$

The interpretation of the above formula is as follows [19]. If a transaction t contains more frequent patterns, its FPOF value will be big, which indicates that it is unlikely to be an outlier.

In contrast, transactions with small FPOF values are likely to be outliers or to be considered as an interesting alert to be investigated by the administrator.

By using $x \subseteq t, X \in FPS \sum_{(D, minisupport)} support(X)$s and re-order the IDS alerts by the simple FPOF for simplicity. Accordingly, we will have the important alerts on the top of IDS log and irrelevant alerts will be pushed to the end of the log file.

**Algorithm 1**. (Alert:Computing Frequent Patterns with Alerts)
Algorithm AlertFp()
Input: A list of k-items, Alert Set of k-Alerts, mini-support s.
Output: A list of frequent patterns Fps and the relative Alert.
Begin
1. Scan the Data Set once to compute
2. Compute frequent pattern F1 from candidate k-itemsets
C1 as F1 = {list of k itemset with Alertslist count ≥ minsupport , Alert1counter}.
3. For Fi < k   i=1 m=0 Counter=0 do
Begin
3.1. If Fi € Alertmi then counter(m)++
3.2. i = i+1, m=m+1
3.3. Compute the next candidate set Ci+1 as F1
End

## 5. CASE STUDY

Snort [22] is a widely used IDS sensor package that applies attack signatures for detecting suspicious network traffic and can emit alerts as syslog messages. Here we will base our case study on the Snort standard attributes (figure1 )

| sid | cid | Sig_id | sig_name | timestamp | ip_src | ip_dst | proto | sport | dport |
|-----|-----|--------|----------|-----------|--------|--------|-------|-------|-------|

**Figure 1 Standard Alert Attributes**

Consider the below Snort sample (figure 2). This sample will be used to simply clarify the idea

```
7 1 508 WEB-MISC/doc/access  25 2 6/11/2010 8:57 AM 1136881320
2148203530 6 46,865 80
7 2 508 WEB-MISC/robots.txt/access 25 2 6/11/2010 8:57 AM 3632363311
2148203629 6 34,074 80
7 3 508 WEB-MISC/robots.txt/access 25 2 8/11/2010 8:59 AM 3632363313
2148203229 6 34,075 80
```

**Figure 2. Snort alerts sample**

We employed the frequent itemset mining for discovering the frequent patterns from the sample IDS log, figure 3.depicts patterns detection from the above snort sample.

```
* * * * *, t1, t2 ,t3 Support: 3 * * * * 25, t1, t2 ,t3
Support: 3 7 * * * (25) , t1, t2 ,t3
Support: 3
(7) * * * (25) 2, t1, t2 ,t3
Support: 3
(7) * 508 * (25) (2), t1, t2 ,t3
Support: 3
(7) * (508) * (25) (2) * * * * 6, t1, t2 ,t3
Support: 3
(7) * (508) * (25) (2) * * * * (6) * 80, t1, t2 ,t3
Support: 3
(7) * (508) WEB-MISC/robots.txt/access (25) (2) * * * * (6) * (80), t1, t2
Support: 2
(7) * (508) * (25) (2) * 8:57AM * * (6) * (80), t1, t2
Support: 2
(7) * (508) * (25) (2) 6/11/2010 (8:57AM) * * (6) * (80) , t1, t2
Support: 2
```

**Figure 3. Sample alert patterns**

Finally the alerts are simple sorted in ascending order according to their weight (simple FPOF) and top p% of them is put into the set of candidate true alerts. Definition 1 (Frequent Pattern Outlier Factor [19]) Let D = {t1,…tn} be a set of n transactions. upport( ) denotes the ratio between the number of transactions that contain itemset X and the number of transactions in D. minisupport is a user defined threshold. It defines the minimal support of frequent itemset. And FPS(D,minisupport) is a set of frequent patterns mined from D with minisupport. Given a transaction t, its Frequent Pattern Outlier Factor (FPOF) is calculated. Finally the alerts are sorted in ascending order according to their weight and top p% of them is put into the set of candidate true alerts.

```
simple FPOF t(3) = 7
7 3 508 WEB-MISC/robots.txt/access 25 2 8/11/2010 8:59AM 3632363313 2148203229 6 34,075 80
simple FPOF t(1) = 8
7 1 508 WEB-MISC/doc/access 25 2 6/11/2010 8:57AM 1136881320 2148203530 6 46,865 80 simple FPOF t(2) = 9
7 2 508 WEB-MISC/robots.txt/access 25 2 6/11/2010 8:57AM 3632363311 2148203629 6 34,074 80
```

**Figure 4. Output sample**

# 6. IMPLEMENTATION AND PERFORMANCE

In this section, we describe our classifier implementation and experiments. In our setup, alerts sorted in a new separate log file for further review. Classifiers are rebuilt every midnight using the IDS sensor log data. Once the frequent pattern has been detected for, it will be used for further alert classification. This allows for the classifier to adapt to new routine alert patterns with a reasonable learning time. The Outlier Factor will be calculated for each transaction, then we will resort the transactions accordingly (figure 4).

> **Attack1**: 9 59,231 916 ARE/BYTE/UNICODE 0 3 6/22/2010 9:09AM 1812014676 2148203529 6 null null
> **Attack2:** 9 25,747 748 FTP/CWD 38 2 6/22/2010 5:31PM 1812014676 2148203530 6 50,74121
> **Attack3:** 9 340,939 991 WEB/MISC/Chunked 31 1 6/22/2010 3:20PM 1812014676 2148203529 6 50,249 80
> **Attack4:** 9 340,950 882 WEB-CGI/cart32.exe/access 25 2 6/22/2010 3:21PM 1812014676 2148203529 6 50,708 80
> **Attack5:** 9 1,529 541 FTP/CWD~attempt 34 2 6/22/2010 11:24AM 1812014676 2148203533 6 60,500 21

**Figure5. attempted attacks from 108.1.38.84**

In our experimental we have applied 5 artificial hacks from a specific source IP to be monitored on our result.
Table 2 presents our experiment results on 22 June 2010 (with 28,670 records) that contains different 5 known attempted attacks (figure 5).

| mini-support | frequent itemsets | attempted 5 attacks place | reduction |
|---|---|---|---|
| 2 | 101101522 | first 7 records | 99.975 % |
| 3 | 45443340 | first 11 records | 99.962 % |
| 4 | 32589268 | first 24 records | 99.916 % |
| 5 | 25684452 | first 28 records | 99.902 % |
| 6 | 23664252 | first 34 records | 99.882 % |
| 7 | 21317488 | first 34 records | 99.882 % |

**Table 2: minisupport/attaches place on the output**

These results are comparable with the performance of other recent data mining methods which have reduced the number of alerts by 74-97% [16, 2, 9, 10, 1].

During the experiments, we have measure the system reliability and accuracy for different support values / frequent itemsets founded these results are comparable with the original attempted attacks and its place on the output file.

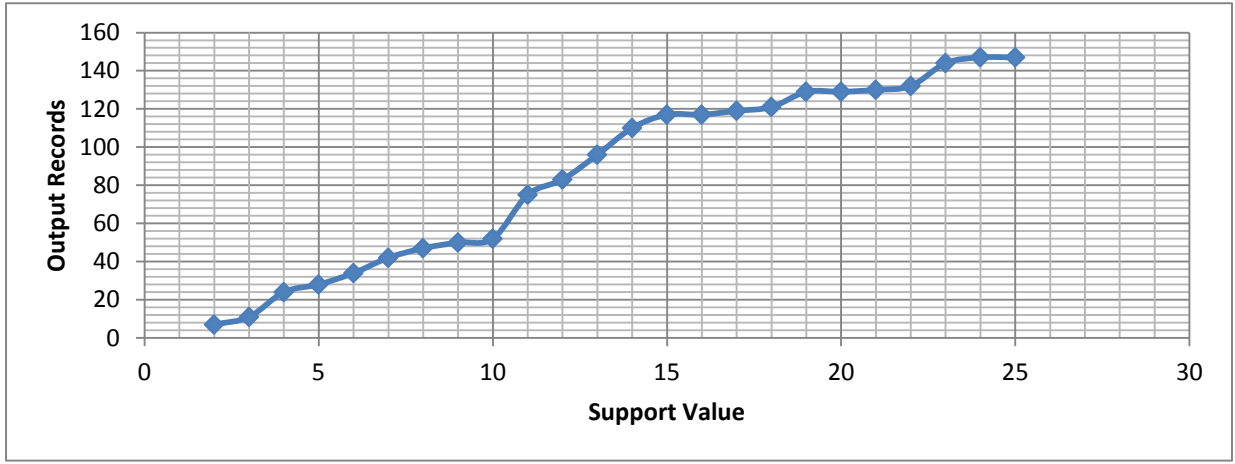

**Figure6. mini support value vs. the 5 attacks in output**

Finally the security analysts will find the IDS logs sorted according to the relevant importance.

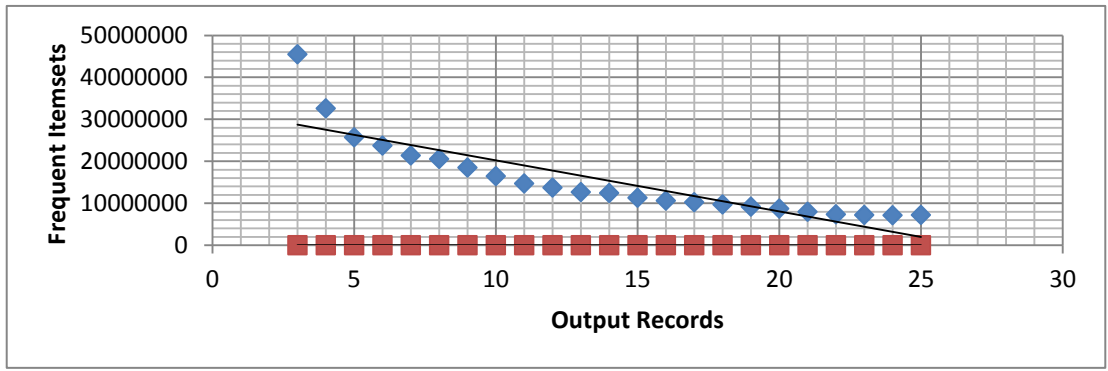

**Figure7. mini support value vs. founded frequent pattern**

## 7. OPEN ISSUES AND FUTURE WORK

In this paper, we have presented a novel data mining based IDS alert classification method and sorted for the security analysts according to the alert importance.
Although our preliminary results are promising, one issue remains open – major changes in the arrival rate of routine alerts might be symptoms of large scale attacks, but are hard to detect. However, this is an inherent weakness of alert classification and sorting systems (e.g., see [8, 5, 6] for a related discussion). For the future work, we plan to research our classification method further, and study various statistical algorithms (e.g., time series analysis) for detecting unexpected fluctuations in the arrival rates of routine alerts.